\documentclass[a4paper,11pt]{article}
\usepackage{jcappub} 
\usepackage[latin9]{inputenc}
\usepackage{textcomp}
\usepackage{amsmath}
\usepackage{amssymb}
\usepackage{color}

\makeatletter

\newcommand{\be}[1]{\begin{equation}\label{#1}}
\newcommand{\ee}{\end{equation}}
\newcommand{\ba}[1]{\begin{eqnarray}\label{#1}}
\newcommand{\ea}{\end{eqnarray}}

\makeatother

\begin{document} 
\title{On the Consistency of the Wheeler-DeWitt Equation in the Quantized Eddington-inspired Born-Infeld Gravity}
\author[a,b]{Mariam Bouhmadi-L\'{o}pez,}
\author[c,d]{Che-Yu Chen}
\author[c,d,e]{and Pisin Chen}
\affiliation[a]{Department of Theoretical Physics, University of the Basque Country
UPV/EHU, P.O. Box 644, 48080 Bilbao, Spain\\}
\affiliation[b]{IKERBASQUE, Basque Foundation for Science, 48011, Bilbao, Spain\\}
\affiliation[c]{Department of Physics and Center for Theoretical Sciences, National Taiwan University, Taipei, Taiwan 10617\\}
\affiliation[d]{LeCosPA, National Taiwan University, Taipei, Taiwan 10617\\}
\affiliation[e]{Kavli Institute for Particle Astrophysics and Cosmology, SLAC National Accelerator Laboratory, Stanford University, Stanford, CA 94305, USA}
\emailAdd{mariam.bouhmadi@ehu.eus}
\emailAdd{b97202056@gmail.com}
\emailAdd{pisinchen@phys.ntu.edu.tw}

\abstract{We re-examine the quantum geometrodynamical approach within the Eddington-inspired-Born-Infeld theory of gravity, which was first proposed in our previous work \cite{Bouhmadi-Lopez:2016dcf}. A thorough analysis of the classical Hamiltonian with constraints is carried and the correctness and self-consistency of the modified Wheeler deWitt equation (WDW) is studied. We find that based on the newly obtained WDW equation derived with the use of the Dirac brackets, the conclusion reached in Ref.~\cite{Bouhmadi-Lopez:2016dcf} can be corroborated. The big rip singularity present in the classical theory, and induced by a phantom perfect fluid, is expected to be avoided when quantum effects encoded on the modified WDW equation are taken into account.}

\maketitle
\flushbottom

\section{Introduction}
\label{sec:intro}
In the context of Einstein's general relativity (GR), the existence of spacetime singularities has been proven to be unavoidable in many different physical configurations. For example, in standard big bang cosmology as well as in various black hole geometries, a spacetime singularity where the underlying theory ceases to be valid is known to be ubiquitous. Furthermore, spacetime singularities not only appear at small scale such as the big bang singularity and the black hole singularity, but may also exist at extremely large scales. One of the example is the big rip singularity associated with certain types of phantom dark energy \cite{Nojiri:2005sx,Starobinsky:1999yw,Caldwell:1999ew,Caldwell:2003vq,Carroll:2003st,Chimento:2003qy,Dabrowski:2003jm,GonzalezDiaz:2003rf,GonzalezDiaz:2004vq,BouhmadiLopez:2004me,BouhmadiLopez:2006fu,Bamba:2012cp,Elizalde:2005ju}. The latter is introduced as a candidate for the explanation of the current accelerated expansion of our universe \cite{Perlmutter:1998np,Riess:1998cb} and it violates the null energy condition. The big rip singularity is actually a generic spacetime singularity driven by phantom dark energy. The size as well as the curvature of the universe diverge at the singularity and in a finite cosmic time. Before reaching the singularity where the spacetime would be destroyed, all bounded structures would be ripped asunder by the strong Hubble flow. 

In Refs.~\cite{Bouhmadi-Lopez:2013lha,Bouhmadi-Lopez:2014jfa}, we have proven that in the Eddington-inspired-Born-Infeld theory of gravity (EiBI) filled with phantom dark energy, the big rip singularity is unavoidable, even though the theory is able to remove the big bang singularity \cite{Banados:2010ix,Scargill:2012kg}. The existence of spacetime singularities indicates that the underlying theory is still incomplete and some quantum effects are expected to come into play near these singularities. In Ref.~\cite{Bouhmadi-Lopez:2016dcf}, we have investigated the possibility of the avoidance of the big rip singularity by considering quantum effects in the EiBI theory. The scrutiny was essentially based on the quantum geometrodynamical approach in which the Wheeler-deWitt (WDW) equation plays a central role in describing the quantum behavior of the universe \cite{qgkiefer}. However, in Ref.~\cite{Bouhmadi-Lopez:2016dcf} the derivation of the WDW equation was not well justified. The importance of the fact that the EiBI theory contains multiple constraint equations was underestimated and a further and deeper analysis is necessary. 

In this paper, we will make up for the aforementioned missing link in deriving the WDW equation. More explicitly, we will $i)$ perform a thorough and complete analysis of the constraints at the classical level, $ii)$ identify the first class as well as the second class constraints in the system, and $iii)$ promote the canonical variables to quantum operators by using the Dirac bracket. After obtaining a self-consistent WDW equation which actually resembles that derived in \cite{Bouhmadi-Lopez:2016dcf}, we will prove that the solution to the WDW equation fulfills the DeWitt boundary condition \cite{DeWitt:1967yk} and the big rip singularity is expected to be avoided by quantum effects. Note that in Ref.~\cite{Albarran:2017swy}, we have followed a preliminary procedure to study the quantum avoidance of other cosmological abrupt events, such as the little rip and the little sibling of big rip, in the EiBI phantom model.  

This paper is outlined as follows. In section~\ref{sec:eibiclassical}, we briefly review the classical EiBI phantom model and exhibit how a big rip singularity would appear within this model. In section~\ref{seclangran}, we construct the modified WDW equation by considering an alternative action proposed in Ref.~\cite{Delsate:2012ky}. The system is shown to be a constrained system and a thorough classical analysis as well as the quantization with the Dirac brackets are performed. In section~\ref{solvewdw}, we solve the WDW equations under two different factor orderings and show that the big rip singularity is hinted to be avoided according to the DeWitt criterion. We finally present our conclusions in section~\ref{sec:conclusion}.

\section{The EiBI phantom model: constant equation of state}
\label{sec:eibiclassical}
The gravitational action of the EiBI theory is given by \cite{Banados:2010ix}
\begin{equation}
S_{EiBI}=\frac{2}{\kappa}\int d^4x\Big[\sqrt{|g_{\mu\nu}+\kappa R_{(\mu\nu)}(\Gamma)|}-\lambda\sqrt{-g}\Big]+S_M(g),
\label{actioneibi}
\end{equation}
where $|g_{\mu\nu}+\kappa R_{(\mu\nu)}(\Gamma)|$ is the absolute value of the determinant of the rank two tensor $g_{\mu\nu}+\kappa R_{(\mu\nu)}(\Gamma)$. The theory is constructed on the Palatini formalism, in which the metric $g_{\mu\nu}$ and the affine connection $\Gamma$ should be treated as independent variables. In addition, $R_{(\mu\nu)}(\Gamma)$ is the symmetric part of the Ricci tensor constructed by $\Gamma$. Furthermore, $S_M$ stands for the matter Lagrangian, where matter is assumed to be coupled only to the metric $g$. Additionally, $\lambda$ is a dimensionless constant quantifying an effective cosmological constant at the low curvature limit. The parameter $\kappa$ is a constant characterizing the theory and we will restrict our analysis to a positive $\kappa$, in order to avoid the instabilities associated with an imaginary effective sound speed usually present in the EiBI theory with a negative $\kappa$ \cite{Avelino:2012ge}. Finally, in this paper we will assume $8\pi G=c=1$.

Since the theory contains two independent variables, $g_{\mu\nu}$ and $\Gamma$, the equations of motion should be obtained by varying the action \eqref{actioneibi} with respect to both of them. The affine structure of the theory implies that the affine connection $\Gamma$ would not be the Christoffel symbols of the physical metric $g_{\mu\nu}$. Instead, it can be proven that there exists a second auxiliary metric $q_{\mu\nu}$ such that it is compatible with the affine connection and defines the curvatures. After varying the action, this auxiliary metric $q_{\mu\nu}$ can be obtained: $\lambda q_{\mu\nu}=g_{\mu\nu}+\kappa R_{(\mu\nu)}(\Gamma)$. 

In our previous works \cite{Bouhmadi-Lopez:2013lha,Bouhmadi-Lopez:2014jfa}, we considered a homogeneous, isotropic and flat universe filled with a phantom dark energy component which is described by a perfect fluid whose equation of state is constant and such that $w<-1$. It was proven that even though the EiBI theory is free of the initial big bang singularity, the big rip singularity is still unavoidable. More precisely, the scale factor $a$ would increase without limits in a finite cosmic time from now and the Hubble rate $H\equiv\dot{a}/a$ as well as its cosmic time derivative can be approximated as follows
\begin{align}
H^2&\approx\frac{4\sqrt{|w|^3}}{3(3w+1)^2}\rho\rightarrow\infty,\nonumber\\
\dot{H}&\approx\frac{2\sqrt{|w|^3}}{(3w+1)^2}|1+w|\rho\rightarrow\infty,
\end{align}
in the far future. On the above equations the dot denotes the derivative with respect to the cosmic time $t$. The phantom energy density $\rho$ and pressure $p=w\rho$ diverge as well at that regime. At the same time, the scale factor $a(t)$ blows up and this cosmic doomsday is what is named as the big rip singularity, which in this case it happens in the EiBI phantom model. 

\section{The modified WDW equation}\label{seclangran}
Since the big rip singularity is unavoidable in the EiBI phantom model, in the following sections we are going to investigate whether the big rip singularity can be smeared by some sorts of quantum effects. We will essentially follow the same strategy employed in Ref.~\cite{Bouhmadi-Lopez:2016dcf} in which a quantum geometrodynamical approach was used. This approach is basically rooted on the canonical quantization aiming towards a quantum theory of gravity. In this framework, the construction of the WDW equation, which results from quantizing the classical Hamiltonian of the theory, plays a crucial role. However, in Ref.~\cite{Bouhmadi-Lopez:2016dcf} the derivation of the modified WDW equation was mostly based on physical intuition, while a complete and consistent mathematical treatment was not complete and even not fully adequate. Therefore, in this paper we will re-examine the derivation of the modified WDW equation in detail and study more carefully the constraints of the system, which is fundamental in this model. We will show that after obtaining the modified WDW equation, one can reach the same physical conclusion as obtained in Ref.~\cite{Bouhmadi-Lopez:2016dcf}, i.e., the wave function vanishes when approaching the singularity and the big rip singularity is expected to be avoided.
 
\subsection{The effective Lagrangian}
 In Refs.~\cite{Bouhmadi-Lopez:2016dcf,Albarran:2017swy}, we have shown that the classical Hamiltonian describing the gravitational system is much easier to obtain via an alternative action \cite{Delsate:2012ky}:
\begin{equation}
S_{a}=\lambda\int d^4x\sqrt{-q}\Big[R(q)-\frac{2\lambda}{\kappa}+\frac{1}{\kappa}(q^{\alpha\beta}g_{\alpha\beta}-2\sqrt{\frac{g}{q}})\Big]+S_M(g),
\label{actionalternative}
\end{equation}
where $R(q)\equiv q^{\alpha\beta}R_{\beta\alpha}(q)$. Note that $q^{\mu\nu}$ stands for the inverse of the auxiliary metric $q_{\mu\nu}$. The action \eqref{actionalternative} is similar to a bi-gravity action without dynamics for $g_{\mu\nu}$ and it is dynamically equivalent to the action \eqref{actioneibi} in the sense that they give the same classical field equations \cite{Delsate:2012ky}. However, the action \eqref{actionalternative} can be seen as the EiBI action written in the Einstein frame because it is linear on $R(q)$, and most importantly it does not contain a square root of the curvature, i.e., a square root involving second order derivatives of the scale factor of the metric compatible with the affine connection $\Gamma$. Therefore, the construction of the classical Hamiltonian is more straightforward starting from the action \eqref{actionalternative}.

Firstly, we assume a homogeneous and isotropic universe:
\begin{align}
g_{\mu\nu}dx^{\mu}dx^{\nu}&=-N(t)^2 dt^2+a(t)^2d\vec{x}^2,\nonumber\\
q_{\mu\nu}dx^{\mu}dx^{\nu}&=-M(t)^2 dt^2+b(t)^2d\vec{x}^2,\nonumber
\end{align}
where $N(t)$ and $M(t)$ are the lapse functions of $g_{\mu\nu}$ and $q_{\mu\nu}$, respectively. Similarly, $a$ and $b$ correspond to the scale factor of each metric. Then, we consider the simplest case in which the matter component is described by a perfect fluid with a given equation of state. Therefore, the matter can be purely described by the scale factor $a$.

After integrations by part, the reduced Lagrangian constructed from the action \eqref{actionalternative}, $S_a=v_0\int dt\mathcal{L}$, can be obtained as
\begin{equation}
\mathcal{L}=\lambda Mb^3\Big[-\frac{6{\dot{b}}^2}{M^2b^2}-\frac{2\lambda}{\kappa}+\frac{1}{\kappa}\Big(\frac{N^2}{M^2}+3\frac{a^2}{b^2}-2\frac{Na^3}{Mb^3}\Big)\Big]-2\rho(a)Na^3,
\label{LA}
\end{equation}
where $v_0$ corresponds to the spatial volume after a proper compactification for spatially flat sections \cite{qgkiefer}. For the sake of convenience, we introduce two changes of variables:
 \begin{equation}
X\equiv\frac{N}{M}\,,\qquad Y\equiv\frac{a}{b}\,,
\end{equation}
and rewrite the Lagrangian \eqref{LA} as follows
\begin{equation}
\mathcal{L}=\lambda Mb^3\Big[-\frac{6\dot{b}^2}{M^2b^2}-\frac{2\lambda}{\kappa}+\frac{1}{\kappa}(X^2+3Y^2-2XY^3)\Big]-2\rho(bY)Mb^3XY^3.
\label{LAA}
\end{equation}
Notice that the energy density $\rho$ is written as a function of $a=bY$ in the Lagrangian.

\subsection{The classical analysis of the Hamiltonian}
As we will next show, the system described by the Lagrangian \eqref{LAA} is a constrained system. The conjugate momenta of the system can be obtained as follows
\begin{align}
p_b\equiv&\frac{\partial\mathcal{L}}{\partial\dot{b}}=-\frac{12\lambda b}{M}\dot{b}\,,\\
p_X\equiv&\frac{\partial\mathcal{L}}{\partial\dot{X}}=0\,,\qquad
p_Y\equiv\frac{\partial\mathcal{L}}{\partial\dot{Y}}=0\,,\qquad
p_M\equiv\frac{\partial\mathcal{L}}{\partial\dot{M}}=0\,.\label{3.6}
\end{align}
According to Eqs.~\eqref{3.6}, the system has three primary constraints \cite{Henneaux,Diraclecture}
\begin{equation}
p_X\sim0\,,\qquad
p_Y\sim0\,,\qquad
p_M\sim0\,,
\end{equation}
where $\sim$ denotes the weak equality, i.e., equality on the constraint surface on which all the constraints in the phase space are satisfied. Given the existence of primary constraints, the total Hamiltonian is defined as follows \cite{Henneaux,Diraclecture}
\begin{align}
\mathcal{H}_T=&\,\dot{b}p_b-\mathcal{L}+\lambda_Xp_X+\lambda_Yp_Y+\lambda_Mp_M\nonumber\\
=&\,M\Big[-\frac{p_b^2}{24\lambda b}+\frac{2\lambda^2b^3}{\kappa}-\frac{\lambda}{\kappa}b^3X^2-\frac{3\lambda}{\kappa}b^3Y^2+\frac{2XY^3b^3}{\kappa}(\lambda+\kappa\rho)\Big]\nonumber\\
&+\lambda_Xp_X+\lambda_Yp_Y+\lambda_Mp_M,
\end{align}
where $\lambda_X$, $\lambda_Y$, and $\lambda_M$ are Lagrangian multipliers associated with each primary constraint. Note that the primary constraints are obtained directly from the definition of the conjugate momenta, i.e., Eqs.~\eqref{3.6}. These constraints should be satisfied throughout time and this would lead to more constraints which are called secondary constraints in the system. Deriving these additional constraints requires the use of the equations of motion: $\dot{\phi}=[\phi,\mathcal{H}_T]\sim0$, where $\phi$ stands for the primary constraints \cite{Henneaux,Diraclecture}. We would name these requirements consistent conditions of the constraints{\footnote{The Poisson bracket is defined as
\begin{equation}
[F,G]=\frac{\partial F}{\partial q_i}\frac{\partial G}{\partial p_i}-\frac{\partial F}{\partial p_i}\frac{\partial G}{\partial q_i},\nonumber
\end{equation}
where $q_i$ are the variables and $p_i$ their conjugate momenta. Notice that repeating suffices denote the summation over them.}}.

According to the consistent conditions of each primary constraint, that is, their conservation in time: $[p_X,\mathcal{H}_T]\sim0$, $[p_Y,\mathcal{H}_T]\sim0$, and $[p_M,\mathcal{H}_T]\sim0$, one can derive three secondary constraints \cite{Henneaux,Diraclecture}:
\begin{align}
C_X&\equiv\lambda X-Y^3(\lambda+\kappa\rho)\sim0,\\
C_Y&\equiv3\lambda-3XY(\lambda+\kappa\rho)-XY^2b\kappa\rho'\sim0,\\
C_M&\equiv\frac{p_b^2}{24\lambda b}-\frac{2\lambda^2b^3}{\kappa}+\frac{\lambda}{\kappa}b^3X^2+\frac{3\lambda}{\kappa}b^3Y^2-\frac{2XY^3b^3}{\kappa}(\lambda+\kappa\rho)\sim0.
\end{align}
The prime denotes the derivative with respect to $a=bY$. Note that these secondary constraints are the Euler-Lagrangian equations of their corresponding variables $X$, $Y$ and $M$, respectively. Furthermore, it can be shown that the total Hamiltonian is also a constraint of the system
\begin{equation}
\mathcal{H}_T=-MC_M+\lambda_Xp_X+\lambda_Yp_Y+\lambda_Mp_M\sim 0\label{htttt}.
\end{equation}
Because the Poisson brackets of the total Hamiltonian with all constraints should vanish weakly by definition, $\mathcal{H}_T$ is a first class constraint and we will use it to construct the modified WDW equation.

Furthermore, the Poisson brackets of the secondary constraints with the total Hamiltonian can be written as
\begin{align}
[C_X,\mathcal{H}_T]&=\frac{\partial C_X}{\partial X}\lambda_X+\frac{\partial C_X}{\partial Y}\lambda_Y+\frac{\partial C_X}{\partial M}\lambda_M+\frac{\partial C_X}{\partial b}\frac{\partial\mathcal{H}_T}{\partial p_b}\nonumber\\
&=\frac{\partial C_X}{\partial X}\lambda_X+\frac{\partial C_X}{\partial Y}\lambda_Y-M\frac{\partial C_X}{\partial b}\frac{\partial C_M}{\partial p_b},\label{n1}\\
[C_Y,\mathcal{H}_T]&=\frac{\partial C_Y}{\partial X}\lambda_X+\frac{\partial C_Y}{\partial Y}\lambda_Y+\frac{\partial C_Y}{\partial M}\lambda_M+\frac{\partial C_Y}{\partial b}\frac{\partial\mathcal{H}_T}{\partial p_b}\nonumber\\
&=\frac{\partial C_Y}{\partial X}\lambda_X+\frac{\partial C_Y}{\partial Y}\lambda_Y-M\frac{\partial C_Y}{\partial b}\frac{\partial C_M}{\partial p_b},\label{a1}\\
[C_M,\mathcal{H}_T]&=\frac{\partial C_M}{\partial X}\lambda_X+\frac{\partial C_M}{\partial Y}\lambda_Y+\frac{\partial C_M}{\partial M}\lambda_M+\frac{\partial C_M}{\partial b}\frac{\partial\mathcal{H}_T}{\partial p_b}-\frac{\partial C_M}{\partial p_b}\frac{\partial\mathcal{H}_T}{\partial b}\nonumber\\
&=\frac{2b^3}{\kappa}(C_X\lambda_X+YC_Y\lambda_Y)\sim0.\label{m1}
\end{align}
Again, the consistent conditions of the secondary constraints require the above Poisson brackets to vanish weakly. It can be easily seen that the following two equations $[C_X,\mathcal{H}_T]\sim0$ and $[C_Y,\mathcal{H}_T]\sim0$ form a coupled equation and the Lagrangian multipliers $\lambda_X$ and $\lambda_Y$ can be determined by solving this coupled equation. On the other hand, the conservation of $C_M$ in time, i.e., $[C_M,\mathcal{H}_T]\sim0$, contributes as an equation of the form $0=0$. Therefore, there is no other independent constraints in the system. Most importantly, it can also be seen that $\lambda_M$ remains undetermined after calculating the consistent conditions of all constraints. This hints at the existence of a first class constraint associated with $\lambda_M$ in the system, as will be shown more explicitly later.

The system has six constraints: $p_X$, $p_Y$, $p_M$, $C_X$, $C_Y$, and $C_M$. We consider the following Poisson brackets among these constraints:
\begin{align}
[C_X,p_X]&=\frac{\partial C_X}{\partial X}=\lambda,\nonumber\\
[C_Y,p_Y]&=\frac{\partial C_Y}{\partial Y}=-3X(\lambda+\kappa\rho)-5XYb\kappa\rho'-XY^2b^2\kappa\rho'',\nonumber\\
[C_X,C_M]&=\frac{\partial C_X}{\partial b}\frac{\partial C_M}{\partial p_b}=-\frac{Y^4\kappa\rho'p_b}{12\lambda b}.
\end{align}
Therefore, the constraints $p_X$, $p_Y$, $C_X$, $C_Y$, and $C_M$ are second class constraints because for each constraint there exists at least one non-vanishing Poisson bracket with other constraints. On the other hand, the primary constraint $p_M$ is found to be a first class constraint because its Poisson brackets with other constraints vanish weakly \cite{Henneaux,Diraclecture}. Recall that the Lagrangian multiplier $\lambda_M$ cannot be determined by calculating the consistent conditions of all the constraints.

The existence of the first class constraint $p_{M}$ implies a gauge degree of freedom in the system. To proceed, we will find an appropriate gauge condition to fix this gauge. An appropriate gauge fixing condition $f$ should satisfy two criteria \cite{Henneaux,Diraclecture}:
\begin{align}
 [f,p_{M}]\nsim&\,0,\\ 
 [f,\mathcal{H}_T]\sim&\,0.\label{3.18cri}
\end{align}
The first criterion means that the gauge condition $f$ compensates the gauge degree of freedom generated by the first class constraint $p_{M}$. In other words, the constraint $p_{M}$ becomes a second class constraint after fixing the gauge because one of its Poisson bracket with the other constraint is not zero, i.e., $[f,p_{M}]\nsim0$. The second criterion, i.e., Eq.~\eqref{3.18cri}, is the consistent condition of this gauge choice. An appropriate choice of $f$ is $f=M$, which means $M$ is a constant. After fixing the gauge, the conservation in time of the gauge condition, i.e., $[M,\mathcal{H}_T]=0$, implies $\lambda_M=0$.

\subsection{Quantization of the system}
As mentioned in the previous subsection, to construct the modified WDW equation we will impose the first class constraint $\mathcal{H}_T$ as a restriction on the Hilbert space on which the wave function of the universe $\left|\Psi\right\rangle$ is defined, $\hat{\mathcal{H}}_T\left|\Psi\right\rangle=0$. The hat denotes the operator. The remaining constraints $\chi_i=\{M,\,p_M,\,p_X,\,p_Y,\,C_X,\,C_Y\}$ are all second class and we need to consider the Dirac brackets to construct the commutation relations and promote the phase space functions to operators \cite{Diraclecture}. Note that $C_M$ can be used to construct the first class constraint $\mathcal{H}_T$ by taking a linear combination of these constraints, i.e., Eq.~\eqref{htttt}, so it should be excluded from the set $\chi_i$.

In general, for a constrained system containing several constraints, some of them first class and some second class, one can apply linear combinations of these constraints to find as many first class constraints as possible. The remaining constraints will be then second class, say $\chi_i$, and one cannot obtain first class constraint anymore by taking linear combinations of them. The Dirac bracket of two phase space functions $F$ and $G$ is then defined by \cite{Diraclecture}
\begin{equation}
[F,G]_D=[F,G]-[F,\chi_i]\Delta_{ij}[\chi_j,G],
\end{equation}
where $\Delta_{ij}$ is a matrix satisfying
\begin{equation}
\Delta_{ij}[\chi_j,\chi_k]=\delta_{ik}.
\end{equation}
In principle, if all first class constraints were exhausted by taking linear combinations of the constraints, the matrix $[\chi_j,\chi_k]$ is invertible and $\Delta_{ij}$ exists. This statement has been proven in Dirac's lecture \cite{Diraclecture}.

The first important property of the Dirac bracket is that the classical equations of motion can be obtained through it:
\begin{align}
[F,\mathcal{H}_T]_D&=[F,\mathcal{H}_T]-[F,\chi_j]\Delta_{jk}[\chi_k,\mathcal{H}_T]\nonumber\\
&\sim[F,\mathcal{H}_T]\nonumber\\
&\sim\dot{F}.
\end{align}
The first weak equality holds because $\mathcal{H}_T$ is first class and we have $[\chi_k,\mathcal{H}_T]\sim0$.
The second crucial property is that the Dirac bracket of a second class constraint with any phase space function is zero strongly (without inserting any constraint):
\begin{align}
[\chi_i,G]_D&=[\chi_i,G]-[\chi_i,\chi_j]\Delta_{jk}[\chi_k,G]\nonumber\\
&=[\chi_i,G]-\delta_{ik}[\chi_k,G]\nonumber\\
&=0.
\end{align}
This property suggests that to quantize a system with second class constraints, it is necessary to use the Dirac brackets to define the commutation relations \cite{Diraclecture}:
\begin{equation}
[\hat{F},\hat{G}]=i\hbar[F,G]_{D,\,(F=\hat{F},\,G=\hat{G})},
\label{commute}
\end{equation}
because we can treat all second class constraints $\chi_i$ in the system as zero operators within a quantum interpretation after defining the commutation relations with the Dirac brackets.

The matrix $\Delta_{ij}$ is the inverse of $[\chi_i,\chi_j]$ and in our system it can be written as follows:
\begin{align}
\Delta_{ij}&=\begin{bmatrix}
[M,M]&[M,p_M]&[M,p_X]&[M,p_Y]&[M,C_X]&[M,C_Y]\\
[p_M,M]&[p_M,p_M]&[p_M,p_X]&[p_M,p_Y]&[p_M,C_X]&[p_M,C_Y]\\
[p_X,M]&[p_X,p_M]&[p_X,p_X]&[p_X,p_Y]&[p_X,C_X]&[p_X,C_Y]\\
[p_Y,M]&[p_Y,p_M]&[p_Y,p_X]&[p_Y,p_Y]&[p_Y,C_X]&[p_Y,C_Y]\\
[C_X,M]&[C_X,p_M]&[C_X,p_X]&[C_X,p_Y]&[C_X,C_X]&[C_X,C_Y]\\
[C_Y,M]&[C_Y,p_M]&[C_Y,p_X]&[C_Y,p_Y]&[C_Y,C_X]&[C_Y,C_Y]
\end{bmatrix}^{-1}\nonumber\\
&=\begin{bmatrix}
0&1&0&0&0&0\\
-1&0&0&0&[p_M,C_X]&[p_M,C_Y]\\
0&0&0&0&[p_X,C_X]&[p_X,C_Y]\\
0&0&0&0&[p_Y,C_X]&[p_Y,C_Y]\\
0&[C_X,p_M]&[C_X,p_X]&[C_X,p_Y]&0&0\\
0&[C_Y,p_M]&[C_Y,p_X]&[C_Y,p_Y]&0&0
\end{bmatrix}^{-1}\nonumber\\
&\sim\begin{bmatrix}
0&1&0&0&0&0\\
-1&0&0&0&0&0\\
0&0&0&0&[p_X,C_X]&[p_X,C_Y]\\
0&0&0&0&[p_Y,C_X]&[p_Y,C_Y]\\
0&0&[C_X,p_X]&[C_X,p_Y]&0&0\\
0&0&[C_Y,p_X]&[C_Y,p_Y]&0&0
\end{bmatrix}^{-1}.
\end{align}
After some calculations, the Dirac brackets between the fundamental variables are
\begin{align}
[b,p_b]_D&=[b,p_b]=1,\nonumber\\
[b,X]_D&=0,\nonumber\\
[b,Y]_D&=0,\nonumber\\
[X,Y]_D&=0,\nonumber\\
[X,p_b]_D&=f_1(X,Y,b)=f_1(b),\nonumber\\
[Y,p_b]_D&=f_2(X,Y,b)=f_2(b),
\label{Dirac}
\end{align}
where $f_1$ and $f_2$ are two non-vanishing functions. Notice that $f_1$ and $f_2$ can be written as functions of $b$ because it is legitimate to insert the constraints $C_X$ and $C_Y$ to replace $X$ and $Y$ with $b$ when calculating the Dirac brackets.

On the $XYb$ basis, we define
\begin{align}
\langle  XYb|\hat{b}|\Psi\rangle&=b\langle XYb|\Psi\rangle,\nonumber\\
\langle XYb|\hat{X}|\Psi\rangle&=X\langle XYb|\Psi\rangle,\nonumber\\
\langle XYb|\hat{Y}|\Psi\rangle&=Y\langle XYb|\Psi\rangle,\nonumber\\
\langle XYb|\hat{p_b}|\Psi\rangle&=-i\hbar\frac{\partial}{\partial b}\langle XYb|\Psi\rangle-i\hbar f_1\frac{\partial}{\partial X}\langle XYb|\Psi\rangle-i\hbar f_2\frac{\partial}{\partial Y}\langle XYb|\Psi\rangle.
\end{align}
It can be checked that the fundamental commutation relations become
\begin{align}
\langle XYb|[\hat{b},\hat{p_b}]|\Psi\rangle&=i\hbar\langle XYb|\Psi\rangle,\nonumber\\
\langle XYb|[\hat{X},\hat{p_b}]|\Psi\rangle&=i\hbar f_1\langle XYb|\Psi\rangle=i\hbar\langle XYb|\hat{f_1}|\Psi\rangle,\nonumber\\
\langle XYb|[\hat{Y},\hat{p_b}]|\Psi\rangle&=i\hbar f_2\langle XYb|\Psi\rangle=i\hbar \langle XYb|\hat{f_2}|\Psi\rangle,\nonumber\\
\langle XYb|[\hat{X},\hat{b}]|\Psi\rangle&=\langle XYb|[\hat{Y},\hat{b}]|\Psi\rangle=\langle XYb|[\hat{X},\hat{Y}]|\Psi\rangle=0,
\end{align}
and therefore they satisfy Eqs.~\eqref{commute} and \eqref{Dirac}. 

To proceed, we perform a redefinition of the wave function $\langle XYb|\rightarrow\langle \xi(X,Y,b), \zeta(X,Y,b), b|$. Based on the chain rules:
\begin{align}
\Big(\frac{\partial}{\partial X}\Big)_{Yb}&=\frac{\partial \xi}{\partial X}\Big(\frac{\partial}{\partial \xi}\Big)_{\zeta b}+\frac{\partial \zeta}{\partial X}\Big(\frac{\partial}{\partial \zeta}\Big)_{\xi b},\nonumber\\
\Big(\frac{\partial}{\partial Y}\Big)_{Xb}&=\frac{\partial \xi}{\partial Y}\Big(\frac{\partial}{\partial \xi}\Big)_{\zeta b}+\frac{\partial \zeta}{\partial Y}\Big(\frac{\partial}{\partial \zeta}\Big)_{\xi b},\nonumber\\
\Big(\frac{\partial}{\partial b}\Big)_{XY}&=\frac{\partial \xi}{\partial b}\Big(\frac{\partial}{\partial \xi}\Big)_{\zeta b}+\frac{\partial \zeta}{\partial b}\Big(\frac{\partial}{\partial \zeta}\Big)_{\xi b}+\Big(\frac{\partial}{\partial b}\Big)_{\xi\zeta},
\end{align}
the momentum operator $\hat{p_b}$ acting on $|\Psi\rangle$ can be expressed on the $\xi\zeta b$ basis as follows:
\begin{align}
&\langle \xi\zeta b|\hat{p_b}|\Psi\rangle\nonumber\\
=&-i\hbar\frac{\partial}{\partial b}\langle XYb|\Psi\rangle-i\hbar f_1\frac{\partial}{\partial X}\langle XYb|\Psi\rangle-i\hbar f_2\frac{\partial}{\partial Y}\langle XYb|\Psi\rangle\nonumber\\
=&-i\hbar\Big(\frac{\partial \xi}{\partial b}\frac{\partial }{\partial \xi}+\frac{\partial \zeta}{\partial b}\frac{\partial }{\partial \zeta}+\frac{\partial }{\partial b}\Big)\langle \xi\zeta b|\Psi\rangle\nonumber\\
&-i\hbar f_1\Big(\frac{\partial \xi}{\partial X}\frac{\partial }{\partial \xi}+\frac{\partial \zeta}{\partial X}\frac{\partial }{\partial \zeta}\Big)\langle \xi\zeta b|\Psi\rangle-i\hbar f_2\Big(\frac{\partial \xi}{\partial Y}\frac{\partial }{\partial \xi}+\frac{\partial \zeta}{\partial Y}\frac{\partial }{\partial \zeta}\Big)\langle \xi\zeta b|\Psi\rangle\nonumber\\
=&-i\hbar\Big(\frac{\partial \xi}{\partial b}+f_1\frac{\partial \xi}{\partial X}+f_2\frac{\partial \xi}{\partial Y}\Big)\frac{\partial}{\partial \xi}\langle \xi\zeta b|\Psi\rangle\nonumber\\
&-i\hbar\Big(\frac{\partial \zeta}{\partial b}+f_1\frac{\partial \zeta}{\partial X}+f_2\frac{\partial \zeta}{\partial Y}\Big)\frac{\partial}{\partial\zeta}\langle\xi\zeta b|\Psi\rangle-i\hbar\frac{\partial}{\partial b}\langle \xi\zeta b|\Psi\rangle.
\label{changeofbasis}
\end{align}
Let us now consider the following first order linear partial differential equation:
\begin{equation}
\frac{\partial W}{\partial b}+f_1\frac{\partial W}{\partial X}+f_2\frac{\partial W}{\partial Y}=0.
\label{PDE}
\end{equation}
This differential equation has two non-degenerate solutions \cite{PDElecture}:
\begin{equation}
W_1=W_1\left[X-\int^bf_1(b')db'\right]\,,\qquad W_2=W_2\left[Y-\int^bf_2(b')db'\right].
\end{equation}
Therefore, if the functions $\xi$ and $\zeta$ are these two non-degenerate solutions respectively, they satisfy Eq.~\eqref{PDE} and the first two terms on the right hand side of the last equality in Eq.~\eqref{changeofbasis} vanish. This means that the expression of $\hat{p_b}$ on the basis $\langle \xi\zeta b|$, such that $\xi$ and $\zeta$ satisfy Eq.~\eqref{PDE}, contains $\partial_b$ only:
\begin{equation}
\langle \xi\zeta b|\hat{p_b}|\Psi\rangle=-i\hbar\frac{\partial}{\partial b}\langle \xi\zeta b|\Psi\rangle.
\label{np}
\end{equation}
Therefore, in the $\xi\zeta b$ basis, the modified WDW equation $\langle\xi\zeta b|\hat{\mathcal{H}}_T|\Psi\rangle=0$ can be rewritten as 
\begin{equation}
\frac{-1}{24\lambda}\langle\xi\zeta b|\frac{\hat{p}_b^2}{b}|\Psi\rangle+V(b)\langle\xi\zeta b|\Psi\rangle=0,
\label{WDWgeneral}
\end{equation}
where the term containing $\hat{p}_b^2$ is determined by Eq.~\eqref{np} and its explicit form depends on the factor orderings. Note that the eigenvalues $X$ and $Y$ can be written as functions of $b$ according to the constraints $C_X$ and $C_Y$, hence it leads to the potential $V(b)$ as follows
\begin{equation}
V(b)=\frac{2\lambda^2b^3}{\kappa}+\frac{\lambda}{\kappa}b^3X^2(b)-\frac{3\lambda}{\kappa}b^3Y^2(b).\label{wdwpotential}
\end{equation}

 \section{Quantum avoidance of the big rip singularity}\label{solvewdw}
The modified WDW equation is shown in Eq.~\eqref{WDWgeneral}, while its explicit form should be determined by the factor ordering. In this section, we will solve the WDW equation at the configuration near the classical big rip singularity and to address a cogent conclusion regarding the singularity avoidance. We will choose two different factor orderings and show that our result is robust against different choices of them. 
\subsection{Factor ordering 1}\label{sub1}
First, we consider $\langle\xi\zeta b|b^3\hat{\mathcal{H}}_T|\Psi\rangle=0$ and choose the following factor ordering
\begin{equation}
b^2\hat{p}_b^2=-\hbar^2\Big(b\frac{\partial}{\partial b}\Big)\Big(b\frac{\partial}{\partial b}\Big)=-\hbar^2\Big(\frac{\partial}{\partial x}\Big)\Big(\frac{\partial}{\partial x}\Big),
\end{equation}
where $x=\ln(\sqrt{\lambda}b)$. Near the big rip singularity, the energy density and pressure of phantom dark energy behave as $\rho\propto a^{\epsilon}$ and $p\propto a^{\epsilon}$, where $\epsilon\equiv-3(1+w)>0$. According to the constraints $C_X$ and $C_Y$, the scale factors $b$ and $a$ are related through $b^4\propto a^{4+2\epsilon}$ asymptotically. Therefore, the modified WDW equation \eqref{WDWgeneral} can be approximated as
\begin{equation}
\Big(\frac{d^2}{dx^2}+\frac{48}{\kappa\hbar^2}\textrm{e}^{6x}\Big)\Psi(x)=0,
\label{diff1}
\end{equation}
when $a$ and $x$ approach infinity. The solution is \cite{mathhandbook}
\begin{equation}
\Psi(x)=C_1J_0(A_1\textrm{e}^{3x})+C_2Y_0(A_1\textrm{e}^{3x}),
\end{equation}
where $C_1$ and $C_2$ are constants. Consequently when $x\rightarrow\infty$, its asymptotic behavior reads \cite{mathhandbook}
\begin{equation}
\Psi(x)\approx\sqrt{\frac{2}{\pi A_1}}\textrm{e}^{-3x/2}\Big[C_1\cos{\Big(A_1\textrm{e}^{3x}-\frac{\pi}{4}\Big)}+C_2\sin{\Big(A_1\textrm{e}^{3x}-\frac{\pi}{4}\Big)}\Big],
\end{equation}
where 
\begin{equation}
A_1\equiv\frac{4}{\sqrt{3\kappa\hbar^2}}.
\end{equation}
Here $J_\nu(x)$ and $Y_\nu(x)$ are Bessel function of the first kind and the second kind, respectively \cite{mathhandbook}. Therefore, the wave function $\Psi(x)$ approaches zero when $a$ as well as $x$ go to infinity and the big rip singularity is expected to be avoided according to the DeWitt criterium \cite{DeWitt:1967yk}.

\subsection{Factor ordering 2}\label{sub2}
Starting from the WDW equation \eqref{WDWgeneral}, we can rewrite it by choosing another factor ordering:
\begin{equation}
\frac{\hat{p}_b^2}{b}=-\hbar^2\Big(\frac{1}{\sqrt{b}}\frac{\partial}{\partial b}\Big)\Big(\frac{1}{\sqrt{b}}\frac{\partial}{\partial b}\Big).
\label{pbb}
\end{equation}
Near the big rip singularity, the modified WDW equation can be approximated as
\begin{equation}
\Big(\frac{d^2}{dy^2}+\frac{64}{3\kappa\hbar^2}y^2\Big)\Psi(y)=0,
\label{diff2}
\end{equation}
where we introduce a new variable $y\equiv(\sqrt{\lambda}b)^{3/2}$. The big rip singularity corresponds to the configuration where $y\rightarrow\infty$. Before proceeding further, we highlight that this quantization is based on the Laplace-Beltrami operator which is the Laplacian operator in minisuperspace \cite{qgkiefer}. This operator depends on the number of degrees of freedom involved. For the case of a single degree of freedom, it can be written as in Eq.~\eqref{pbb} (c.f. for example \cite{Albarran:2015tga}). 

The solution of the modified WDW equation \eqref{diff2} reads \cite{mathhandbook}
\begin{equation}
\Psi(y)=C_3\sqrt{y}J_{1/4}(A_1y^2)+C_4\sqrt{y}Y_{1/4}(A_1y^2),
\end{equation}
where $C_3$ and $C_4$ are constants. When $y\rightarrow\infty$, the solution becomes
\begin{equation}
\Psi(y)\approx\sqrt{\frac{2}{\pi A_1y}}\Big[C_3\cos{\Big(A_1y^2-\frac{3\pi}{8}\Big)}+C_4\sin{\Big(A_1y^2-\frac{3\pi}{8}\Big)}\Big].
\end{equation}
Therefore, the wave function $\Psi(y)$ approaches zero when $a$ as well as $y$ go to infinity. According to the DeWitt criterium for singularity avoidance \cite{DeWitt:1967yk}, the big rip singularity is expected to be avoided in this case. In addition, given the results shown in subsections \ref{sub1} and \ref{sub2}, the quantum avoidance of the big rip singularity seem to be independent of the choice of the factor orderings.

\section{Conclusion}\label{sec:conclusion}
In our previous paper \cite{Bouhmadi-Lopez:2016dcf}, we studied the quantum geometrodynamics in the context of the EiBI theory to see whether the big rip singularity can be avoided when quantum effects are taken into account. The WDW equation plays a central role in this approach because it is essentially the WDW equation that describes the quantum behavior of the whole universe in this setup.

In order to construct the WDW equation, one needs to derive a correct and self-consistent Hamiltonian at the classical level. In practice, the Hamiltonian is a first class constraint in the system and it is regarded as a restriction in the Hilbert space to obtain the WDW equation at the quantum level. In that paper \cite{Bouhmadi-Lopez:2016dcf}, we made up some not fully consistent links in the derivation of the WDW equation. A complete and correct constraint analysis is now carried out in this paper. Due to the affine structure in the EiBI theory, the system contains additional auxiliary fields but in the end the system turns out to be reducible because of the second class constraints. 

We identify all the first class and second class constraints in the system. One of the first class constraint $p_M$ corresponds to a gauge degree of freedom and it can be fixed by adding one more additional constraint (fixing the gauge). The other first class constraint, which is the total Hamiltonian, is used to derive the WDW equation. On the other hand, the existence of second class constraints in the system implies the need to use the Dirac brackets when promoting the canonical variables to quantum operators. We have shown that with a complete and self-consistent treatment of the quantization, the WDW equation can be derived and it can also be significantly simplified (cf. Eqs.~\eqref{WDWgeneral} and \eqref{wdwpotential}). 

Finally, we choose two different factor orderings to solve the WDW equation in a configuration near the classical big rip. The solutions that we have found satisfy the DeWitt boundary conditions in the sense that they vanish near the classical big rip singularity. Therefore, we conclude that the big rip singularity in the EiBI phantom model is hinted to be avoidable when quantum effects are considered, and this result seems to be independent of the choice of the factor orderings.

\acknowledgments

The work of MBL is supported by the Basque Foundation of Science Ikerbasque. She also would like to acknowledge the partial support from the Basque government Grant No. IT956-16 (Spain) and from the project FIS2017-85076-P (MINECO/AEI/FEDER, UE). CYC and PC are supported by Taiwan National Science Council under Project No. NSC 97-2112-M-002-026-MY3, Leung Center for Cosmology and Particle Astrophysics (LeCosPA) of National Taiwan University, and Taiwan National Center for Theoretical
Sciences (NCTS). PC is in addition supported by US Department of Energy under Contract No. DE-AC03-76SF00515. CYC would like to thank the department of Theoretical Physics and History of Science of the University of the Basque Country (UPV/EHU) for kind hospitality while part of this work was done.


\begin{thebibliography}{99}

\bibitem{Bouhmadi-Lopez:2016dcf}
  M.~Bouhmadi-L\'opez and C.~Y.~Chen,
  \emph{Towards the Quantization of Eddington-inspired-Born-Infeld Theory},
  \emph{JCAP} {\bf 1611} (2016) no.11,  023
  [\href{http://arxiv.org/abs/1609.00700}{arXiv:1609.00700}].

\bibitem{Starobinsky:1999yw} 
  A.~A.~Starobinsky,
  \emph{Future and origin of our universe: Modern view},
  \emph{Grav.\ Cosmol.\ } {\bf 6} (2000) 157 [\href{http://arxiv.org/abs/astro-ph/9912054}{astro-ph/9912054}].

\bibitem{Caldwell:1999ew} 
  R.~R.~Caldwell,
  \emph{A Phantom menace?},
  \emph{Phys.\ Lett.\ B} {\bf 545} (2002) 23 [\href{http://arxiv.org/abs/astro-ph/9908168}{astro-ph/9908168}].

\bibitem{Carroll:2003st} 
  S.~M.~Carroll, M.~Hoffman and M.~Trodden,
  \emph{Can the dark energy equation - of - state parameter w be less than -1?},
  \emph{Phys.\ Rev.\ D} {\bf 68} (2003) 023509 [\href{http://arxiv.org/abs/astro-ph/0301273}{astro-ph/0301273}].

\bibitem{Caldwell:2003vq}
  R.~R.~Caldwell, M.~Kamionkowski and N.~N.~Weinberg,
  \emph{Phantom energy and cosmic doomsday},
  \emph{Phys.\ Rev.\ Lett.\ } {\bf 91} (2003) 071301 [\href{http://arxiv.org/abs/astro-ph/0302506}{astro-ph/0302506}].

\bibitem{Chimento:2003qy} 
  L.~P.~Chimento and R.~Lazkoz,
  \emph{On the link between phantom and standard cosmologies},
  \emph{Phys.\ Rev.\ Lett.\ } {\bf 91} (2003) 211301 [\href{http://arxiv.org/abs/gr-qc/0307111}{gr-qc/0307111}].

\bibitem{Dabrowski:2003jm} 
  M.~P.~D\c{a}browski, T.~Stachowiak and M.~Szyd{\l }owski,
  \emph{Phantom cosmologies},
  \emph{Phys.\ Rev.\ D} {\bf 68} (2003) 103519 [\href{http://arxiv.org/abs/hep-th/0307128}{hep-th/0307128}].

\bibitem{GonzalezDiaz:2003rf} 
  P.~F.~Gonz\'{a}lez-D\'{i}az,
  \emph{K-essential phantom energy: Doomsday around the corner?},
  \emph{Phys.\ Lett.\ B} {\bf 586} (2004) 1 [\href{http://arxiv.org/abs/astro-ph/0312579}{astro-ph/0312579}].

\bibitem{GonzalezDiaz:2004vq} 
  P.~F.~Gonz\'{a}lez-D\'{i}az,
  \emph{Axion phantom energy},
  \emph{Phys.\ Rev.\ D} {\bf 69} (2004) 063522 [\href{http://arxiv.org/abs/hep-th/0401082}{hep-th/0401082}].

\bibitem{BouhmadiLopez:2004me}
  M.~Bouhmadi-L\'{o}pez and J.~A.~Jim\'{e}nez Madrid,
  \emph{Escaping the big rip?},
  \emph{JCAP} {\bf 0505} (2005) 005 [\href{http://arxiv.org/abs/astro-ph/0404540}{astro-ph/0404540}].

\bibitem{Nojiri:2005sx} 
  S.~Nojiri, S.~D.~Odintsov and S.~Tsujikawa,
  \emph{Properties of singularities in (phantom) dark energy universe},
  \emph{Phys.\ Rev.\ D} {\bf 71} (2005) 063004 [\href{http://arxiv.org/abs/hep-th/0501025}{hep-th/0501025}].

\bibitem{Elizalde:2005ju}
  E.~Elizalde, S.~Nojiri, S.~D.~Odintsov and P.~Wang,
  \emph{Dark energy: Vacuum fluctuations, the effective phantom phase, and holography},
  \emph{Phys.\ Rev.\ D} {\bf 71} (2005) 103504 [\href{http://arxiv.org/abs/hep-th/0502082}{hep-th/0502082}].

\bibitem{BouhmadiLopez:2006fu} 
  M.~Bouhmadi-L\'{o}pez, P.~F.~Gonz\'{a}lez-D\'{i}az and P.~Mart\'{i}n-Moruno,
  \emph{Worse than a big rip?},
  \emph{Phys.\ Lett.\ B} {\bf 659} (2008) 1 [\href{http://arxiv.org/abs/gr-qc/0612135}{gr-qc/0612135}].

\bibitem{Bamba:2012cp}
  K.~Bamba, S.~Capozziello, S.~Nojiri and S.~D.~Odintsov,
  \emph{Dark energy cosmology: the equivalent description via different theoretical models and cosmography tests},
  \emph{Astrophys.\ Space Sci.\ } {\bf 342} (2012) 155 [\href{http://arxiv.org/abs/1205.3421}{arXiv:1205.3421}].

\bibitem{Riess:1998cb} 
  A.~G.~Riess {\it et al.} [Supernova Search Team Collaboration],
  \emph{Observational evidence from supernovae for an accelerating universe and a cosmological constant},
  \emph{Astron.\ J.\ } {\bf 116} (1998) 1009 [\href{http://arxiv.org/abs/astro-ph/9805201}{astro-ph/9805201}].

\bibitem{Perlmutter:1998np} 
  S.~Perlmutter {\it et al.} [Supernova Cosmology Project Collaboration],
  \emph{Measurements of Omega and Lambda from 42 high redshift supernovae},
  \emph{Astrophys.\ J.\ } {\bf 517} (1999) 565 [\href{http://arxiv.org/abs/astro-ph/9812133}{astro-ph/9812133}].

\bibitem{Bouhmadi-Lopez:2013lha} 
  M.~Bouhmadi-L\'{o}pez, C.~-Y.~Chen and P.~Chen,
  \emph{Is Eddington-Born-Infeld theory really free of cosmological singularities?},
  \emph{Eur.\ Phys.\ J.\ C} {\bf 74} (2014) 2802 [\href{http://arxiv.org/abs/1302.5013}{arXiv:1302.5013}].

\bibitem{Bouhmadi-Lopez:2014jfa} 
  M.~Bouhmadi-L\'{o}pez, C.~Y.~Chen and P.~Chen,
  \emph{Eddington-Born-Infeld cosmology: a cosmographic approach, a tale of doomsdays and the fate of bound structures},
  \emph{Eur.\ Phys.\ J.\ C} {\bf 75} (2015) no. 2, 90 [\href{http://arxiv.org/abs/1406.6157}{arXiv:1406.6157}].

\bibitem{Banados:2010ix} 
  M.~Ba\~{n}ados and P.~G.~Ferreira,
  \emph{Eddington's theory of gravity and its progeny},
  \emph{Phys.\ Rev.\ Lett.\ } {\bf 105} (2010) 011101 [\href{http://arxiv.org/abs/1006.1769v2}{arXiv:1006.1769v2}]
  Erratum: [\emph{Phys.\ Rev.\ Lett.\ } {\bf 113} (2014) no. 11, 119901].

\bibitem{Scargill:2012kg} 
  J.~H.~C.~Scargill, M.~Ba\~{n}ados and P.~G.~Ferreira,
  \emph{Cosmology with Eddington-inspired Gravity},
  \emph{Phys.\ Rev.\ D} {\bf 86} (2012) 103533 [\href{http://arxiv.org/abs/1210.1521}{arXiv:1210.1521}].

\bibitem{qgkiefer}
C.~Kiefer, \emph{Quantum Gravity}. Second edition (Oxford University Press, Oxford, 2007).

\bibitem{DeWitt:1967yk} 
  B.~S.~DeWitt,
  \emph{Quantum Theory of Gravity. 1. The Canonical Theory},
  \emph{Phys.\ Rev.\ } {\bf 160} (1967) 1113.

\bibitem{Albarran:2017swy}
  I.~Albarran, M.~Bouhmadi-L\'opez, C.~Y.~Chen and P.~Chen,
  \emph{Doomsdays in a modified theory of gravity: A classical and a quantum approach},
  \emph{Phys.\ Lett.\ B} {\bf 772} (2017) 814
  [\href{http://arxiv.org/abs/1703.09263}{arXiv:1703.09263}].

\bibitem{Delsate:2012ky} 
  T.~Delsate and J.~Steinhoff, \emph{New insights on the matter-gravity coupling paradigm}, 
  \emph{Phys.\ Rev.\ Lett.\ } {\bf 109} (2012) 021101
  [\href{http://arxiv.org/abs/1201.4989v3}{arXiv:1201.4989v3}].

\bibitem{Avelino:2012ge} 
  P.~P.~Avelino,
  \emph{Eddington-inspired Born-Infeld gravity: astrophysical and cosmological constraints},
  \emph{Phys.\ Rev.\ D} {\bf 85} (2012) 104053 [\href{http://arxiv.org/abs/1201.2544}{arXiv:1201.2544}].

\bibitem{Diraclecture}
  P.~A.~M.~Dirac, \emph{Lectures on Quantum Mechanics}, Yeshiva University, New York (1964).

\bibitem{Henneaux}
M.~Henneaux and C.~Teitelboim, \emph{Quantization of gauge systems}, Princeton University Press (1992).

\bibitem{PDElecture}
E.~Miersemann, \emph{Partial Differential Equations}, CreateSpace Independent Publishing Platform (2014).

\bibitem{mathhandbook}
M.~Abramowitz and I.~Stegun, \emph{Handbook on Mathematical Functions} (Dover, 1980).  

\bibitem{Albarran:2015tga}
  I.~Albarran and M.~Bouhmadi-L\'opez,
  \emph{Quantisation of the holographic Ricci dark energy model},
  \emph{JCAP} {\bf 1508} (2015) no.08,  051 [\href{http://arxiv.org/abs/1505.01353v2}{arXiv:1505.01353v2}].

\end{thebibliography}
\end{document}